**Title:** A Feasibility Study of Whole-Cerebrum Task-Based fMRI at 0.55 T: Motor and Visual Tasks Using Single-Shot EPI

**Authors:** Parsa Razmara[1], Takfarinas Medani[1], Anand A. Joshi[1], Majid Abbasi Sisara[2], Ye Tian[1], Sophia X. Cui[3], Justin P. Haldar[1], Krishna S. Nayak[1], and Richard M. Leahy[1]

[1] Ming Hsieh Department of Electrical and Computer Engineering, University of Southern California, Los Angeles, CA, United States
[2] Division of Biokinesiology and Physical Therapy, University of Southern California, Los Angeles, CA, United States
[3] Siemens Healthineers, Los Angeles, CA, United States

## Abstract

**Purpose:** To evaluate the feasibility of whole-cerebrum task-based fMRI on a modern 0.55T system, examining the effects of extended scan duration and within-session consistency on functional mapping outcomes and comparison to acquisition with matched tasks at 3.0T.

**Methods:** Six healthy participants were scanned at 0.55T using a single-shot EPI sequence during bilateral finger-tapping and visual checkerboard tasks. Data were processed using an open-source pipeline. To investigate sensitivity improvements, two consecutive 5-minute runs were concatenated. Within-session consistency was also assessed. A representative 3T dataset provided additional reference context for temporal dynamics and statistical sensitivity.

**Results:** Task-evoked activations were detected at 0.55T using Family-Wise Error (FWE) rate correction ($p < 0.05$) in both motor and visual cortices. While single 5-minute runs provided baseline detectability, extending the acquisition to 10 minutes increased t-statistic magnitudes and cluster extents while substantially reducing background noise (47% reduction in voxelwise dispersion). Run-to-run comparisons showed consistent peak locations and statistical significance across participants. Comparison with the 3T reference demonstrated that statistical significance increases with acquisition time so that a 0.55T scan approximately matches 3T significance levels at twice the scan duration across the range compared.

**Conclusion:** Whole-cerebrum task fMRI is feasible at 0.55T using standard EPI and modern high-performance gradients. These results indicate that modest extensions in scan duration can compensate for reduced intrinsic BOLD sensitivity at lower field strengths.

## Introduction

Functional MRI (fMRI) was first introduced more than three decades ago (1) and has since become one of the most important modern tools for noninvasively evaluating human brain function. While there have been historical fMRI studies that used lower-field MRI systems ($\leq$ 1 T), it has been understood for a long time that fMRI experiments using blood oxygenation dependent (BOLD) contrast are generally improved by using higher-field magnets (2–5). This has driven modern BOLD fMRI studies to primarily rely on field strengths $\geq$ 3T. However, although higher fields can offer substantial benefits to the resolution, sensitivity, and specificity of fMRI, there are also some trade-offs, such as reduced compatibility for subjects with implants (6), increasingly severe artifacts when fMRI is combined with complementary functional imaging modalities such as electroencephalography (EEG) (7–9), and increased sensitivity to artifacts arising from inhomogeneities in the B0 magnetic field (10).

In this work, we investigate the performance of a 0.55T MRI system for task-based fMRI applications. Our interest in this topic is fueled by potential future applications, such as simultaneous fMRI-EEG scans (11), where the lower magnetic field strength may reduce mutual interference, although this remains to be systematically quantified (12), and for studies of subjects with neurostimulators that may not be compatible with higher field strengths. Our study is enabled by the fact that our institution has recently installed a 0.55T whole-body human MRI system, reflecting the recent resurgence of interest in lower-field systems for general MRI applications (6, 13–18). Compared to MRI systems with similar field strengths used in historical fMRI studies, this scanner incorporates modern magnet design and a higher-performance gradient system.

Our results are complementary to recent fMRI studies on modern lower-field MRI scanners (19, 20), although our exploration focuses on different questions in different scenarios. In particular, Wang et al. (19) were focused on comparing EPI against bSSFP on a visual task using a protocol that only captured a part of the visual cortex, while Halder et al. (20) focused specifically on the general question of echo-time selection for T2* contrast efficiency without attempting to assess brain function. In this work, we perform task-based fMRI using motor and visual tasks, employing a protocol that achieves whole-cerebrum FOV coverage, with an emphasis on quantifying activation strengths, understanding the impact of using a longer total acquisition, and assessing activation consistency across runs. A preliminary account of portions of this work was presented at a recent conference (21).

We employ a single-shot EPI acquisition and an open-source pipeline to evaluate the feasibility of whole-cerebrum functional mapping. Furthermore, we provide a representative comparison to 3T data to estimate the scan duration required to match t-statistic magnitudes at 0.55T.

## Methods

### *Experimental Design and Data Acquisition*

Six healthy participants (5 male, 1 female; mean age 29.3 ± 4.8 years, range 24–34 years) were enrolled. All participants performed both the finger-tapping and visual stimulation tasks. The study was approved by the Institutional Review Board (IRB) of the University of Southern California. Written informed consent was obtained from all participants prior to experimentation, and none had a history of neurological or psychiatric disorders.

All participants were scanned on a prototype 0.55T scanner (MAGNETOM Aera; Siemens Healthineers, Erlangen, Germany) equipped with "Aera XQ" high-performance shielded gradients (45 mT/m amplitude, 200 T/m/s slew rate) (6) and a 16-channel head-neck receive array. At 0.55T, anatomical imaging was acquired using a T1-weighted MPRAGE sequence (22, 23). Functional data were collected using a single-shot EPI BOLD sequence covering the entire cerebrum, with TE set to 85 ms to match the apparent T2* of gray matter at 0.55T (6, 19, 24). No multiband acceleration was used at either field. Acquisition parameters for both field strengths are summarized in Table 1.

To provide a qualitative reference for spatial localization and statistical magnitude (t-statistics) at conventional field strengths, anatomical and functional images were acquired from one participant using a 3T system (MAGNETOM Prismafit, Siemens Healthineers). Both used the 3T center's standard neuroscience protocols, with the functional scan following the same task paradigms as the 0.55T sessions, as described below. Parallel imaging was not used at 0.55T, where the lower baseline SNR does not favor the SNR penalty of acceleration. Differences in echo time, echo spacing, bandwidth, and total number of slices between the two protocols reflect both hardware performance, with stronger gradient amplitude at 3T, and field-dependent BOLD-contrast considerations.

The experimental design included two block-design tasks, finger-tapping and visual checkerboard (timing and volume counts in Table 1). The finger-tapping task used self-paced bilateral tapping of all fingers of both hands (25), and the visual task presented a high-contrast flickering checkerboard (26). A single run of each of the two tasks on one of the subjects was also collected on the 3T scanner as described above.

### *Data Analysis*

Functional data were preprocessed in MATLAB (The MathWorks, Natick, MA, USA) using BrainSuite fMRI Pipeline (BFP) (27, 28) for the correlation analysis and SPM12 (29) for the GLM, with identical parameters across both. The preprocessing steps included slice-timing correction and deobliquing. Motion correction used each tool's default: Structural Similarity Index Measure (SSIM)-referenced alignment (BFP) and rigid-body realignment (SPM12). The reference was selected via SSIM optimization, where pairwise SSIM was computed between each frame and every 10th volume; the frame with the highest mean SSIM was chosen as the target. Spatial smoothing was applied using an 8 mm FWHM Gaussian kernel. A temporal high-pass filtering (0.005 Hz cutoff) was used to remove low-frequency drift. Functional volumes were co-registered to subject-specific T1-weighted images via a rigid transformation and non-linearly transformed to the MNI space using the unified segmentation algorithm in SPM12 (30) to facilitate comparison with standard higher-field literature.

Activation maps were generated using both voxel-wise general linear model (GLM) and voxel-wise correlation analysis (31): the GLM provides statistical inference (t-values and corrected cluster statistics), while the correlation analysis quantifies temporal fit and non-target signal characteristics. Task conditions were modeled as boxcar functions convolved with the SPM canonical hemodynamic response function (HRF) (29) to generate a reference hemodynamic waveform. To investigate the effect of extended scan durations, we concatenated two consecutive runs, doubling the number of time points to enhance activation significance (32). For the GLM analysis, statistical significance was assessed using t-tests, and results were thresholded with family-wise error (FWE) correction ($p < 0.05$), with concatenated peaks additionally surviving $p < 0.001$. Cluster extent was computed in SPM12 as the number of contiguous suprathreshold voxels above the FWE-corrected threshold. All statistics are within-subject (first-level), computed across time points. No group-level analysis was performed. For visualization of activation extent, 0.55T activation maps were displayed using false discovery rate (FDR) correction at $q < 0.05$, with inference and visualization thresholds reported separately (33, 34). For 0.55T-versus-3T comparisons, the same FWE correction is applied to both fields, so they are compared on equal terms. Maps were visualized using Nilearn (35). For correlation analysis, the Pearson correlation coefficient was calculated between each voxel's preprocessed time series and the reference hemodynamic waveform. As a representative case, non-target correlation characteristics in the visual experiment were summarized by three metrics: $\mu$ (mean correlation in non-target tissue outside the visual cortex), $\sigma$ (voxelwise standard deviation of correlations), and TV (spatial total variation of the correlation map, where lower values indicate a spatially smoother map with fewer sharp spatial transitions).

To evaluate the temporal dynamics of the BOLD signal, single-vertex time courses were extracted from the motor and visual cortices. To quantify the temporal evolution of statistical sensitivity as a function of scan duration, a cumulative GLM analysis was computed on the averaged signal from a 10-vertex neighborhood centered on the peak motor vertex. For progressively increasing truncation lengths, the GLM t-statistic was recomputed. The design matrix included the HRF-convolved task paradigm, an intercept, and a linear drift regressor to account for low-frequency baseline drift. Model parameters were estimated via Ordinary Least Squares (OLS). To determine the acquisition duration required for the 0.55T system to approach the statistical significance observed at 3T, the resulting p-value curves were interpolated in log-space to compute the average duration ratio. This analysis was performed for all 0.55T participants and repeated for the single participant scanned at 3T.

# Results

Figure 1 presents the time series and statistical sensitivity of the BOLD response at 3T and 0.55T. At both field strengths the time series reflect the expected block-design modulation in motor and visual cortices, yielding correlations of 0.77 (motor) and 0.72 (visual) at 0.55T and 0.91 (motor) and 0.77 (visual) at 3T (Figure 1A–B). Analysis of cumulative p-values as a function of scan duration revealed an increasing trend in statistical significance at 0.55T (Figure 1C). These curves indicate that the 0.55T system requires approximately twice the scan duration to reach p-value levels comparable to those at 3T. The single-subject result is largely consistent with a multi-subject analysis of the 0.55T participants (group mean t-curve across 4 participants, 8 runs total; Figure S1), which also indicates a matching duration ratio of approximately 2 (individual-run p-value curves, Figure S2). These values are preliminary empirical estimates from a small cohort rather than definitive scaling factors.

In one of the six participants, the finger-tapping data were excluded after the subject was observed not following the task instructions during the scan. With a single 5-minute run of the bilateral finger-tapping task at 0.55 T, significant activation (FWE-corrected $p < 0.05$) was consistently observed bilaterally in the primary motor cortex (M1) and supplementary motor area (SMA) (Table S1). These single-run acquisitions yielded a mean peak t-score (averaging the two hemispheric peaks per participant) of 9.56 ± 2.44 and mean cluster size of 366.80 ± 394.10 voxels. Concatenating two runs (10 min total) substantially increased these metrics: the mean peak t-score rose to 12.68 ± 3.11 (33% increase), with a corresponding mean cluster size of 797 ± 695 voxels (>2-fold increase). All reported peaks in the concatenated analysis were significant at $p < 0.001$ (FWE-corrected) (Table S2). Representative axial t-statistic maps (Figure 2A) and supplementary coronal slices (Figure S3) illustrate this bilateral activation.

With a single 5-minute visual stimulation run at 0.55 T, significant activation ($p < 0.05$, FWE-corrected) was consistently detected in the primary visual cortex (V1) and adjacent occipital cortex (Table S3), yielding a mean peak t-score of 7.18 ± 0.83 and cluster size of 68.08 ± 50.27 voxels. Concatenating runs to 10 minutes drove a substantial gain in sensitivity: the mean peak t-score increased to 9.24 ± 1.57 (29% increase), and the mean cluster size expanded to 239.83 ± 212.75 voxels (approximately 3.5-fold increase) (Table S4). All reported peaks in the concatenated maps were significant at $p < 0.001$ (FWE-corrected). Representative axial t-statistic maps (Figure 2B) and supplementary axial slices (Figure S4) illustrate this activation. An extended scan duration also yielded substantial improvements in non-target signal characteristics. Concatenation reduced the mean non-target correlation ($\mu$) from 0.169 ± 0.026 to 0.072 ± 0.018 (57% reduction) and voxelwise dispersion ($\sigma$) from 0.094 ± 0.028 to 0.050 ± 0.009 (47% reduction). Spatial total variation (TV) dropped from 8.210 ± 2.577 to 4.746 ± 0.803 (42% reduction) (Figure S5 and Table S5). These shifts indicate a transition toward smoother, more spatially specific activation maps with reduced background spurious activation when scan duration is increased from 5 to 10 minutes.

To assess within-session consistency, we compared two consecutive runs per participant. Significant clusters, peak t-scores, corrected p-values, and MNI coordinates showed consistent correspondence for both motor and visual tasks, as shown in Tables S1 and S3. In comparison with the 3T reference, the same participant scanned at 0.55T showed activation locations and cluster patterns consistent in terms of spatial specificity and statistical significance (Figures 3–**4**, Figure S6). While the 3T maps exhibited larger cluster extents and higher peak t-statistics as expected, the 10-minute 0.55T acquisitions identified the same activation regions (M1/SMA and V1) as the 5-minute 3T reference ($p < 0.001$, FWE-corrected; see captions for peak coordinates and voxel counts).

## Discussion

This study provides a practical demonstration of whole-cerebrum task-based fMRI at 0.55T using a block design with contemporary acquisition and analysis workflows. By evaluating bilateral finger-tapping and visual checkerboard paradigms, we demonstrate that significant task-evoked activations are detectable in whole-cerebrum acquisitions at 0.55T using FWE correction, even within a single 5-minute acquisition window. These baseline results demonstrate that the intrinsic BOLD mechanism is feasible at 0.55T for functional mapping, at least for elucidation of primary sensory cortices.

The comparison of temporal dynamics at 0.55T and 3T (Figure 1) quantifies the sensitivity difference. While the 0.55T time series yield lower correlation coefficients than 3T, they consistently reflect the task-modulated hemodynamic response (Figure 1A-B). The cumulative sensitivity analysis (Figure 1C) indicates that the 0.55T system requires approximately 2× longer acquisition to reach statistical significance levels comparable to 3T, a ratio consistent across the representative single-subject curve and the multi-subject average (Figure S1). This finding reflects the expected sensitivity penalty at lower field strengths (6, 36), but demonstrates that statistical sensitivity accumulates progressively with scan duration, following the expected $\sqrt{N}$ trend. Local deviations from this trend are attributable to the block-design structure. This overall increasing trend supports the strategy of using extended scanning protocols to compensate for reduced intrinsic BOLD contrast at 0.55T.

To contextualize the 0.55T findings, we compared results from a representative participant scanned at both 0.55T (10 min) and 3T (5 min) using identical experimental paradigms. These qualitative and descriptive quantitative comparisons are contextual and are not intended as equivalence tests. Rather, they situate the 0.55T findings within the broader literature and emphasize that, with appropriate timing and modest extensions in scan duration, task-evoked activations at lower field strengths can align with conventional-field expectations in location and statistical significance.

As a practical note for study design at 0.55T, TE and TR were selected from literature-guided defaults rather than optimized settings. TE was matched to the longer gray-matter T2* at 0.55T, near 86 ms, supporting our selection of TE ≈ 85 ms (6, 19, 24, 25, 37, 38). Shorter TE reduces dropout but increases large-vessel weighting, while longer TE increases thermal noise (36, 39, 40). Reduced susceptibility artifacts at 0.55T lessen the need for short TE (6). TR = 3 s balanced T1 recovery, spatial coverage, and scan time given the limited baseline SNR at 0.55T (36, 40, 41). Future work should systematically evaluate TE/TR trade-offs and alternative readouts such as SMS or 3D-EPI at 0.55T (42, 43).

Single-shot EPI was used for its robustness. Alternative sequences such as multi-echo EPI, bSSFP, or 3D-EPI merit further investigation at lower field strengths.

Together, these results indicate that single-shot EPI, paired with an open-source processing pipeline, can yield statistically significant and consistent whole-cerebrum task activations at 0.55T. Given the cohort size and the single-subject 3T reference, these findings should be interpreted as feasibility evidence rather than definitive benchmarks. The code and data will be released to support replication and future work on additional tasks, populations (44), and reconstruction methods tuned for lower-field physiology. Additional near-term directions include machine-learning–based super-resolution to enhance spatial detail (45) and automated artifact detection to improve image quality and consistency (46). While these findings suggest a promising path for 0.55T task fMRI, larger follow-on studies are required to establish generalizability and to position this platform within research and applications such as EEG compatibility and implant-compatible imaging. This aligns with growing interest in lower-field MRI utility (11, 47).

Several limitations should be noted. The cohort is small ($n = 6$ at 0.55T, with one excluded from the finger-tapping analysis due to task non-compliance), and the 3T comparison is based on a single participant. The duration-ratio estimates should therefore be interpreted as preliminary empirical values rather than calibrated benchmarks. Acquisition parameters were drawn from literature-guided defaults rather than formally optimized at 0.55T, and systematic TE/TR optimization is deferred to future work. The tasks evaluated (bilateral finger-tapping, full-field checkerboard) elicit strong primary-cortex BOLD responses, and sensitivity characterization for higher-order or weaker-response paradigms at 0.55T remains an open question.

## Conclusion

This work provides proof-of-concept evidence that whole-cerebrum task-based fMRI at 0.55T is feasible using a standard single-shot EPI sequence and an open-source processing pipeline. Statistically significant ($p < 0.05$, FWE-corrected) and spatially consistent activations were obtained for motor and visual tasks with whole-cerebrum coverage. While higher fields offer greater intrinsic sensitivity, these results show that modest extensions in scan duration can compensate for reduced intrinsic BOLD contrast at lower field strengths. This may be particularly relevant for applications that prioritize EEG compatibility, reduced susceptibility artifacts, or imaging in subjects with higher-field-incompatible implants.

Future directions include evaluating tasks with subtler BOLD responses, particularly in frontal and temporal regions where the reduced geometric distortion of 0.55T may provide a unique advantage. Other important next steps are exploring event-related designs and systematic TE/TR optimization. Resting-state functional connectivity is a natural next direction at 0.55T.

## Acknowledgements

The authors received grant support from the National Institutes of Health (R01EB026299, U01EB023820, R21-HL159533, R01-AR078912, U01-HL167613), National Science Foundation (No. 1828736), and received research support from Siemens Healthineers. We thank Dr. Jiancheng Zhuang (USC Dornsife Neuroimaging Center) for technical support with the 3T MRI acquisitions, and Mary Yung for research coordination.

## Code and Data Availability

The code used for fMRI analysis and anatomical labeling in this study, built on the BrainSuite fMRI Pipeline (BFP), together with the associated data, will be made publicly available upon publication.

# Figures and Tables

**Table 1.** Acquisition parameters at 0.55T and 3T for the anatomical (MPRAGE) and functional (single-shot EPI) protocols. Both block-design tasks began with an initial 6 s rest period (2 volumes). The finger-tapping task (102 volumes, 5.1 min) consisted of five cycles of alternating activation and rest (30 s / 10 volumes each). Activation blocks involved self-paced bilateral tapping of all fingers of both hands, while rest blocks involved no movement with visual fixation on a central white cross on a black screen. Participants were instructed via visual cues to begin and end tapping. The visual checkerboard task (98 volumes, 4.9 min) comprised six cycles of alternating stimulation and rest (24 s / 8 volumes each). Stimulation blocks presented a high-contrast black-and-white flickering checkerboard (12 Hz), while rest blocks displayed a white fixation cross. Participants were instructed to maintain central fixation throughout the run to minimize eye movements. Two consecutive runs per task were acquired at 0.55T.

| Parameter | 0.55 T | 3 T |
|---|---|---|
| **Anatomical: T1-weighted MPRAGE** | | |
| TR / TE / TI | 1500 / 3.08 / 722 ms | 2400 / 2.22 / 1000 ms |
| Flip angle | 20° | 8° |
| Resolution | 1 × 1 × 1.1 $mm^3$ | 0.8 mm isotropic |
| Field of view | 256 × 256 × 263 $mm^3$ | 256 × 240 × 166 $mm^3$ |
| Bandwidth | 150 Hz/pixel | 220 Hz/pixel |
| Acquisition time | 5 min 45 s | 6 min 38 s |
| Parallel imaging | None | None |
| **Functional: single-shot EPI** | | |
| TR / TE | 3000 / 85 ms | 3000 / 30 ms |
| Flip angle | 90° | 80° |
| Field of view | 210 × 170 $mm^2$ | 240 × 240 $mm^2$ |
| Matrix | 64 × 52 | 86 × 86 |
| In-plane resolution | 3.3 × 3.3 $mm^2$ | 2.8 × 2.8 $mm^2$ |
| Slices | 21 axial, 4 mm, 0.8 mm gap (interleaved) | 44 axial, 3 mm |
| Bandwidth | 1565 Hz/pixel | 2528 Hz/pixel |
| Echo spacing | 0.77 ms | 0.54 ms |
| Parallel imaging / multiband | None / none | GRAPPA R = 2 / none |
| Volumes per run (tapping / visual) | 102 / 98 (5.1 / 4.9 min) | 102 / 98 (5.1 / 4.9 min) |
| Runs per task | 2 | 1 |
| **Hardware** | | |
| Gradient amplitude / slew rate | 45 mT/m / 200 T/m/s | 80 mT/m / 200 T/m/s |
| Receive coil | 16-channel head-neck | 20-channel head-neck |

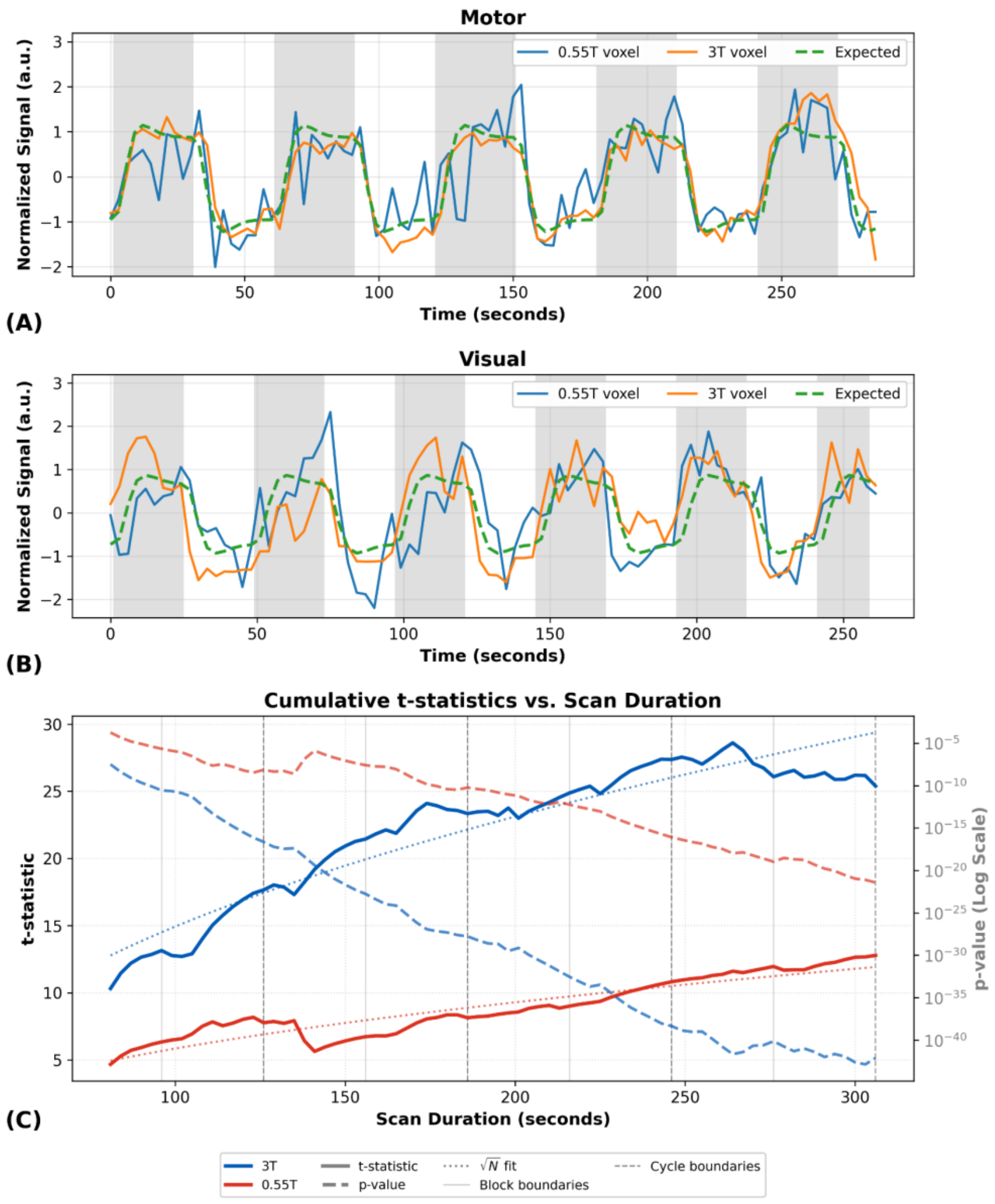

**Figure 1**. BOLD signal time series and statistical sensitivity. (**A-B**) Representative single-vertex task fMRI time series at 0.55T and 3T from a single participant. The plotted vertex was selected based on the maximum correlation with the HRF-convolved reference waveform in the 0.55T data within (**A**) the motor cortex ROI and (**B**) the V1 ROI. The corresponding anatomical location was extracted for 3T comparison. To facilitate visual comparison of temporal dynamics between field strengths, time series were z-score normalized and overlaid with the reference waveform (dashed line). Shaded regions mark task blocks. Pearson correlation (r) between the vertex time series and the reference waveform: (**A**) 0.55T = 0.77, 3T = 0.91; (**B**) 0.55T = 0.72, 3T = 0.77. (**C**) Cumulative t-statistics vs. scan duration for the finger-tapping task. T-values were computed from the signal averaged across a 10-vertex neighborhood centered on the peak motor vertex displayed in (A). Cumulative t-statistics (solid lines, left axis) and corresponding p-values (dashed lines, log-scale right axis) were calculated for the 3T (blue) and 0.55T (red) systems using a General Linear Model (GLM) with linear detrending. Dotted lines show fitted $\sqrt{N}$ envelopes ($t = a\sqrt{N} + b$), reflecting the expected growth of the t-statistic as the standard error of the GLM estimate decreases with increasing sample size (N, the number of time points, proportional to scan

duration). Local fluctuations around this trend correspond to block boundaries in the task design (gray vertical lines mark task-rest transitions; black dashed vertical lines mark complete on-off cycle boundaries), where truncation within an incomplete block temporarily reduces the t-statistic due to imbalanced task and rest samples. While the 3T data reach high statistical significance more rapidly, the 0.55T data exhibit an overall increase in statistical sensitivity with scan duration, requiring approximately 2× longer acquisition to reach p-value levels comparable to 3T.

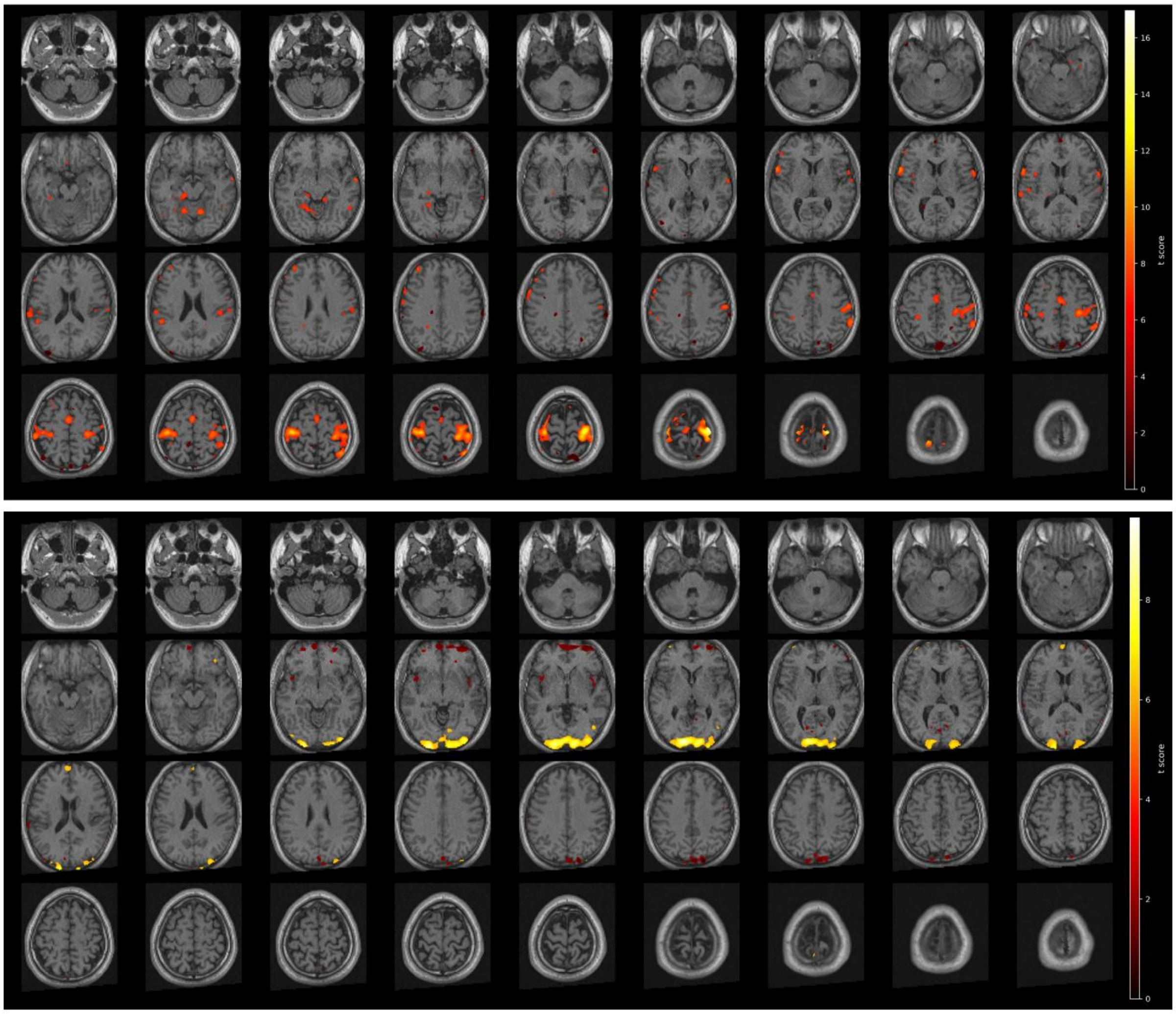

(A)

(B)

**Figure 2.** Statistical activation maps for the finger-tapping and visual tasks at 0.55T. Representative whole-cerebrum t-statistic maps from a single participant, derived from concatenated 10-minute runs and overlaid on the 0.55T T1-weighted anatomical image. (A) Finger-tapping task, showing significant activation in the primary motor cortex; subject-specific statistics (peak t-scores, p-values, MNI coordinates, and cluster sizes) are in Table S2, and selected coronal slices in Figure S3. (B) Visual task, showing significant contiguous activation in the primary visual cortex (V1) and adjacent occipital regions; subject-specific statistics (peak t-scores, p-values, MNI coordinates, and cluster sizes) are in Table S4, and selected axial slices in Figure S4. Maps are displayed at a threshold of $q < 0.05$ (FDR-corrected) to visualize the spatial extent of activation.

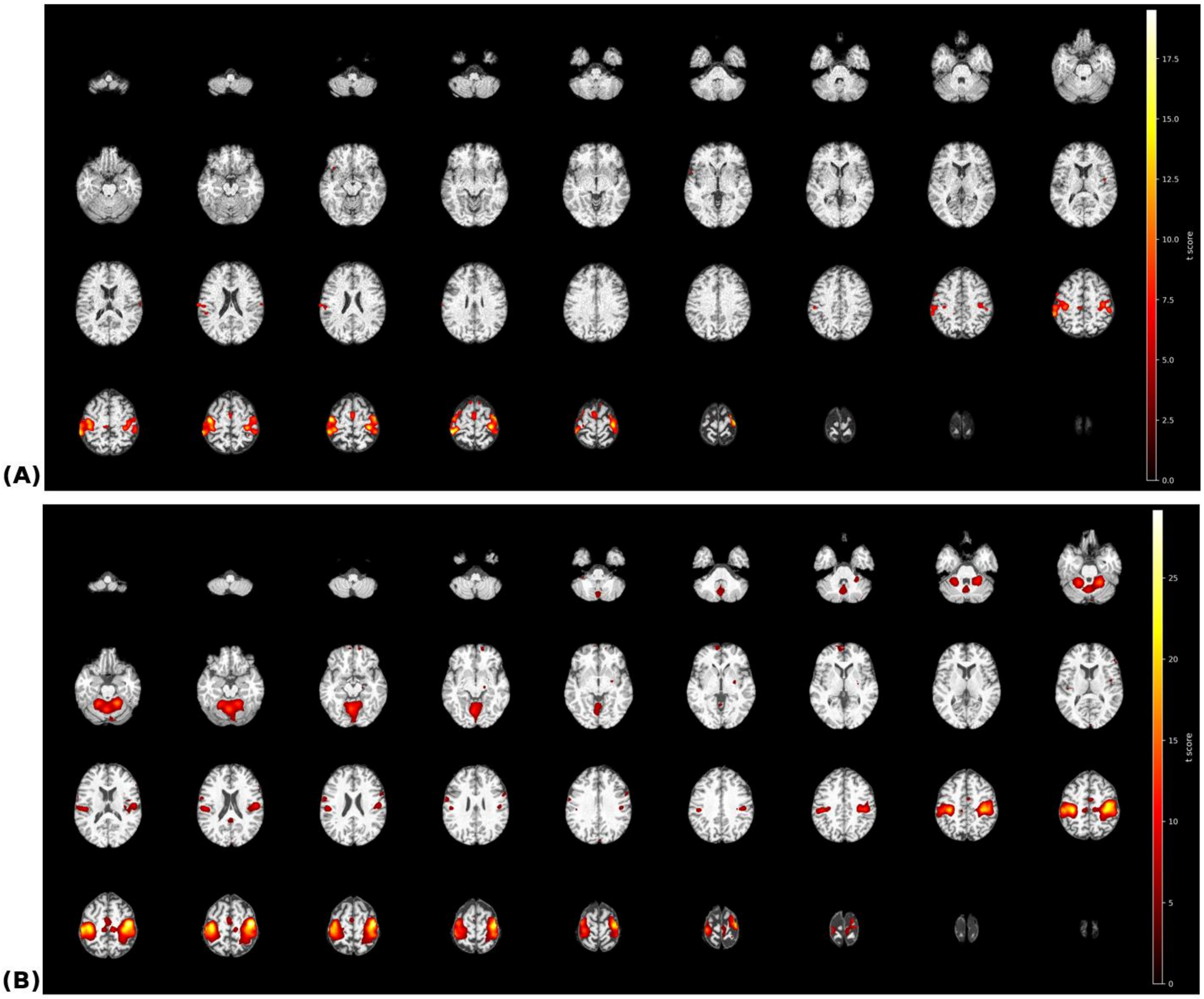

**Figure 3.** 0.55T and 3T fMRI BOLD activation results for a single participant (Motor Task), overlaid on the T1-weighted anatomical images acquired at the respective field strengths. (**A**) t-statistic maps for the finger-tapping experiment acquired at 0.55T. Peak statistics include t=19.2 (MNI: [−44,−10,58], cluster size: 2,070 voxels) and t=16.8 (MNI: [34,−26,62], cluster size: 2,107 voxels). (**B**) Corresponding t-statistic maps acquired at 3T, provided as a reference to evaluate spatial localization. Peak statistics at 3T include t=28.4 (MNI: [42,−14,53], cluster size: 2,013 voxels) and t=24.9 (MNI: [−34,−26,47], cluster size: 1,473 voxels). All reported peaks are significant at $p < 0.001$ (FWE-corrected). Maps are thresholded at $p < 0.05$ (FWE-corrected) and displayed at consistent slice locations, with 3T results presented as a conventional reference to aid in the interpretation of the 0.55T activation locations and significance.

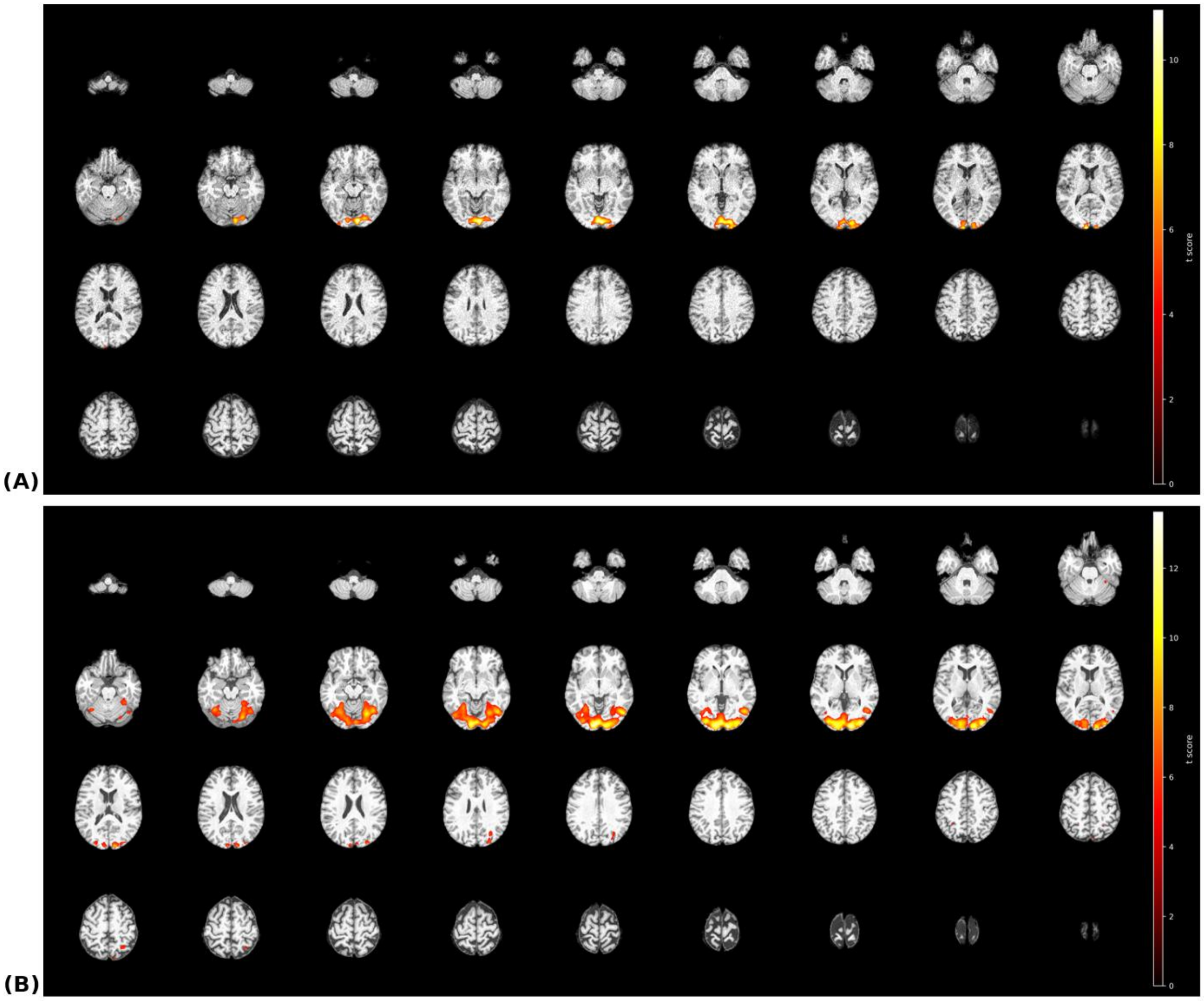

**Figure 4.** 0.55T and 3T fMRI BOLD activation results for a single participant (Visual Task), overlaid on the T1-weighted anatomical images acquired at the respective field strengths. (**A**) t-statistic maps for the visual experiment acquired at 0.55T. Peak statistics include t=11.0 (MNI: [6, -89, -8], cluster size: 302 voxels). (**B**) Corresponding t-statistic maps acquired at 3T. Peak statistics at 3T include t=13.4 (MNI: [-11, -92, -7], cluster size: 2,781 voxels). All reported peaks are significant at $p < 0.001$ (FWE-corrected). Maps are thresholded at $p < 0.05$ (FWE-corrected) and displayed at consistent slice locations, with 3T results presented as a conventional reference to aid in the interpretation of the 0.55T activation locations and significance.

## Supplementary Material

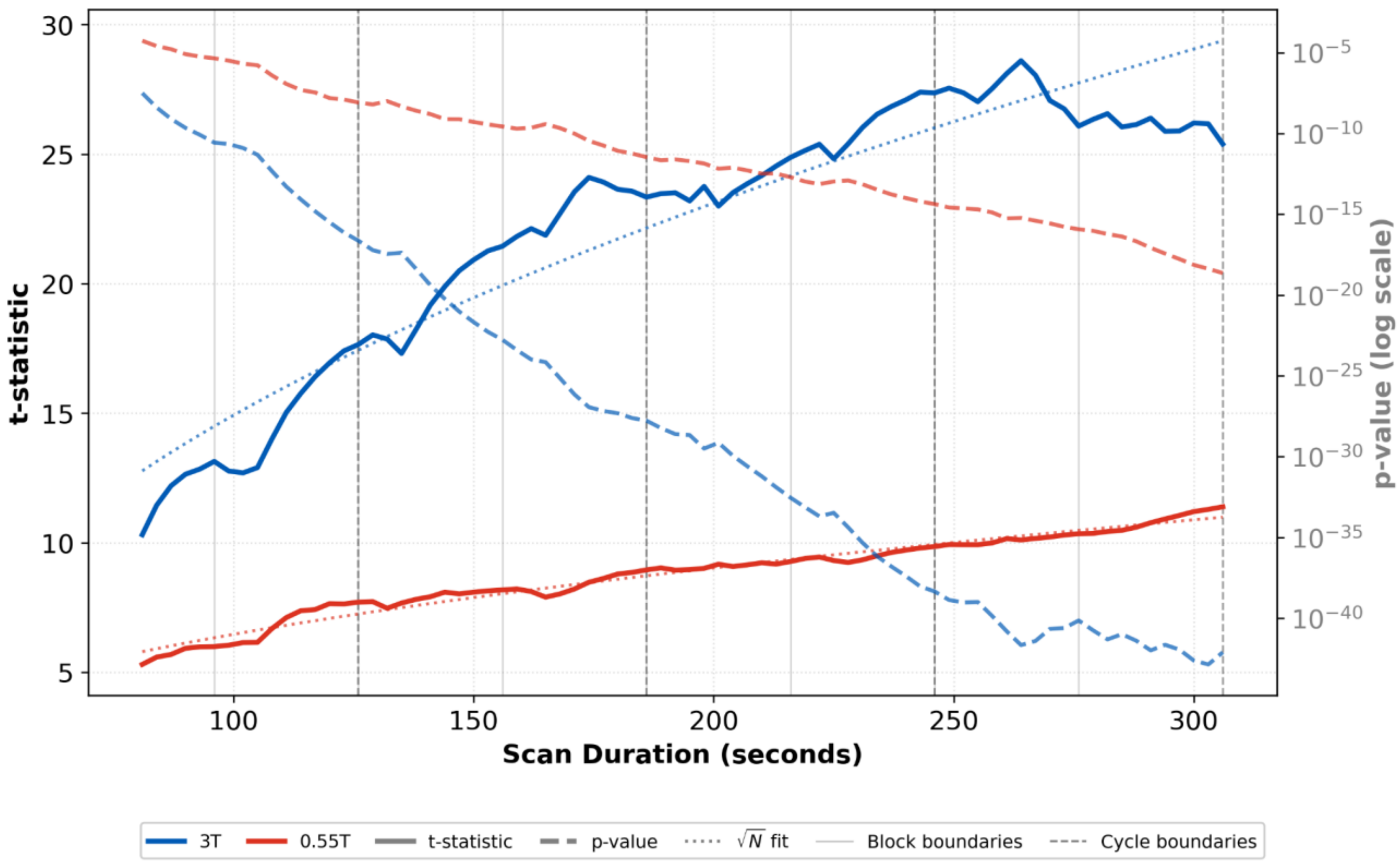

**Supporting Information Figure S1.** Multi-subject cumulative t-statistics vs. scan duration for the finger-tapping task (10-vertex neighborhood). The 0.55T curve (red, solid) represents the mean across four participants (two runs per participant averaged); one participant was excluded from this mean due to weak task activation ($t \approx 5$ at 306 s versus 8–13 for the other four participants). The 3T curve (blue, solid) is from a single representative participant. At each truncation length, t-statistics were computed using a GLM with linear detrending, applied to the averaged signal from a 10-vertex neighborhood centered on the peak motor vertex for each participant. Corresponding p-values are shown as dashed lines (log-scale right axis): for 0.55T, computed from the group-level mean t-statistic (p(mean t)); for 3T, computed directly from the single-participant t-statistic. Dotted lines show fitted √N envelopes ($t = a\sqrt{N} + b$), reflecting the expected growth of the t-statistic as the standard error decreases with increasing sample size. Vertical lines mark task-rest transitions (gray, solid) and complete on-off cycle boundaries (black, dashed). The multi-subject analysis yields a duration ratio of approximately 2× longer acquisition at 0.55T to reach p-value levels comparable to 3T, consistent with the single-subject estimate reported in Figure 1C. Individual-run p-value curves are shown in Figure S2.

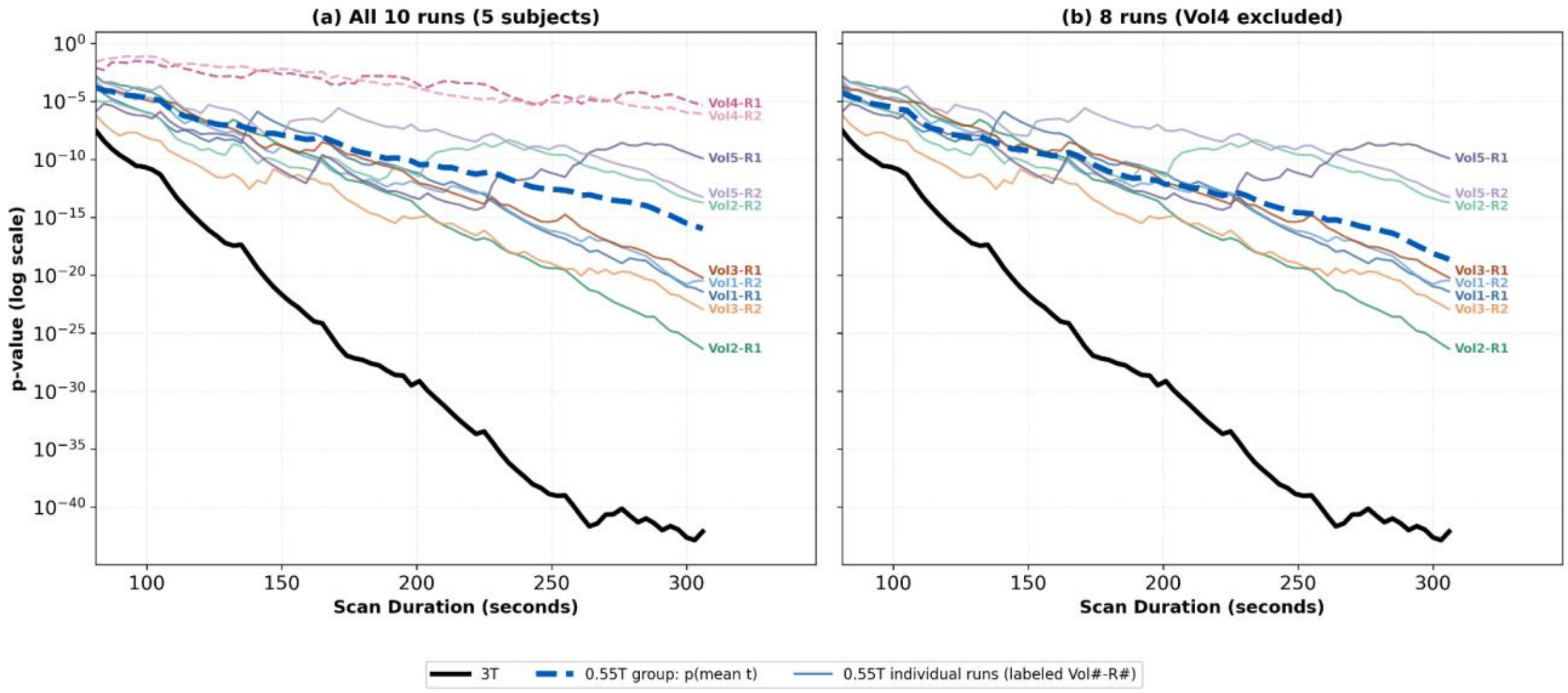

**Supporting Information Figure S2.** Individual-run cumulative p-value curves at 0.55T versus 3T. Ten runs were acquired from five participants (two runs each: Vol1-R1/R2, Vol2-R1/R2, Vol3-R1/R2, Vol4-R1/R2, Vol5-R1/R2). (a) All 10 runs, with Vol4's two runs drawn in dashed pink as the outlier subject identified in Figure S1. (b) Eight runs after excluding Vol4. In both panels, thin colored curves show individual runs (grouped by subject with two shades per subject), the blue dashed curve shows the group-level p-value computed from the mean t-statistic across runs (p(mean t)), and the black curve shows the 3T reference from a single representative participant. The duration ratio computed from the group curves indicates that 0.55T requires approximately 2× longer scan duration than 3T to reach comparable p-value levels, in both panels, demonstrating that this ratio is robust to the inclusion of the outlier participant.

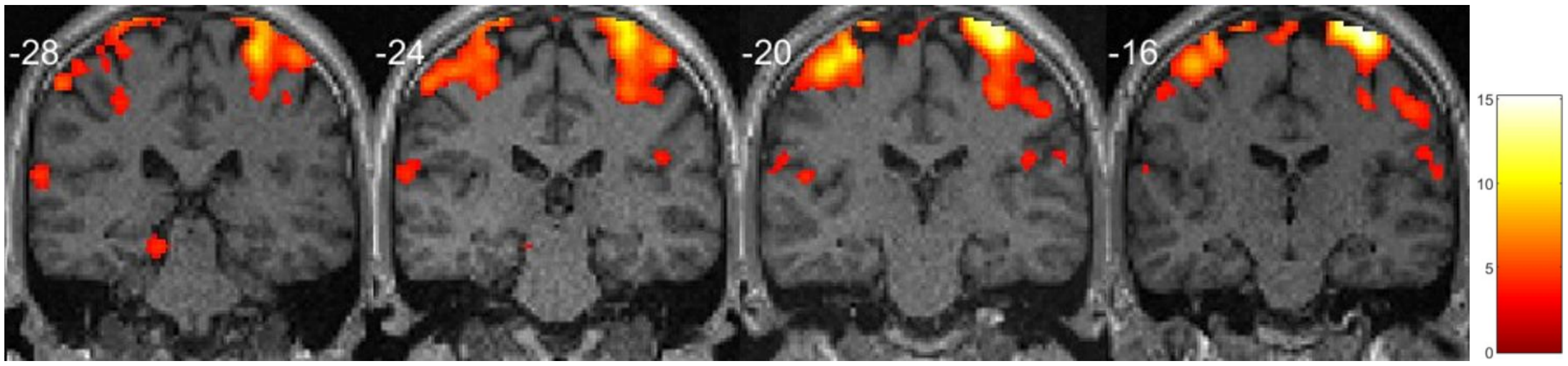

**Supporting Information Figure S3.** Coronal views of motor cortex activation at 0.55T. Statistical activation maps for the bilateral finger-tapping task (same participant as in Figure 2). Coronal slices are displayed at MNI coordinates y = -28, -24, -20, and -16 mm, with t-statistic maps overlaid on the T1-weighted anatomy acquired at 0.55T, highlighting activation in the primary motor cortex. The display threshold matches the main figure: $q < 0.05$ (FDR-corrected).

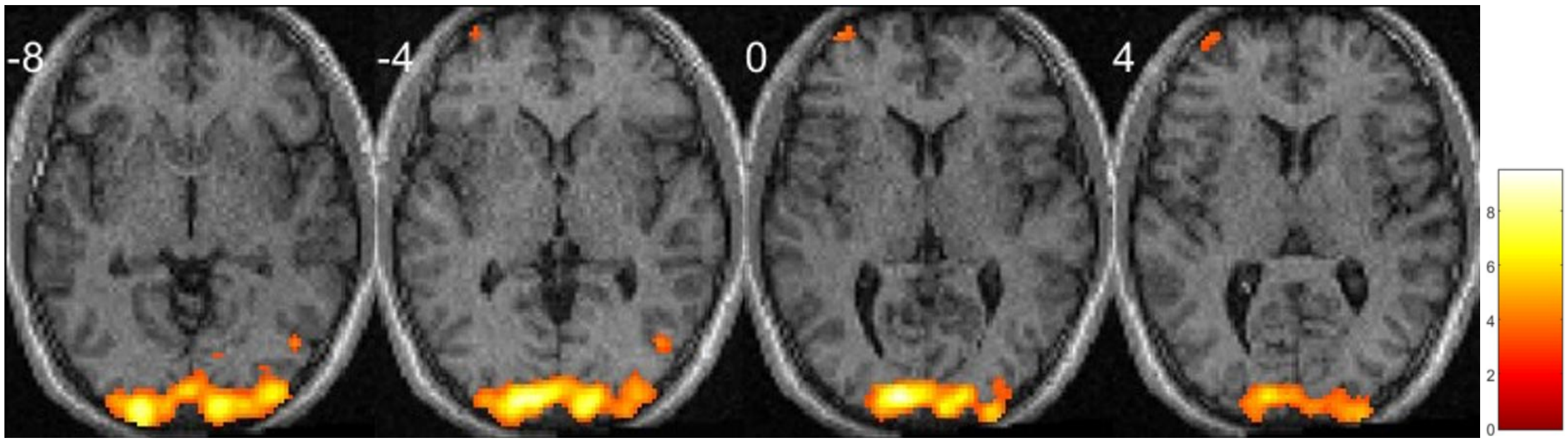

**Supporting Information Figure S4.** Axial views of visual cortex activation at 0.55T. Statistical activation maps for the visual task (same participant as in Figure 2). Axial slices are displayed at MNI coordinates z = -8, -4, 0, and 4 mm, with t-statistic maps overlaid on the T1-weighted anatomy acquired at 0.55T, highlighting activation in the primary visual cortex. The display threshold matches the main figure: $q < 0.05$ (FDR-corrected).

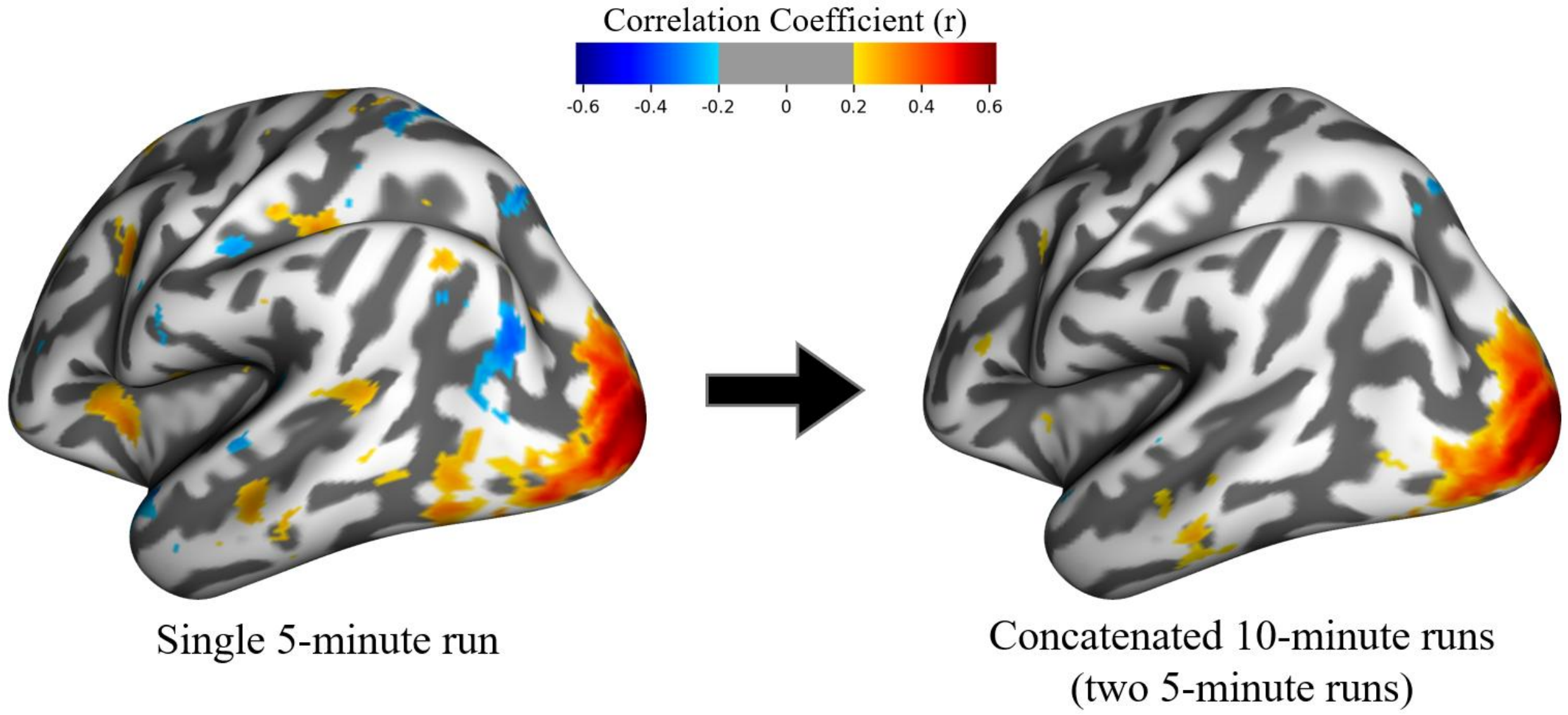

**Supporting Information Figure S5.** Effect of longer scan time on non-target correlation at 0.55T (Visual Task). Extending scan duration by concatenating two 5-minute runs to 10 minutes reduced scattered spurious correlation in non-target areas. Representative correlation surface maps from a single participant display the spatial distribution of Pearson correlation coefficients (r). This reduction is visible in the 10-minute map (right) relative to the single run (left), resulting in improved spatial homogeneity. This qualitative effect is consistent with the quantitative reductions in σ and total variation: see Table S5 for per-participant quantitative metrics (μ, σ, TV). Vertices with $|r| \leq 0.2$ are shown in grayscale.

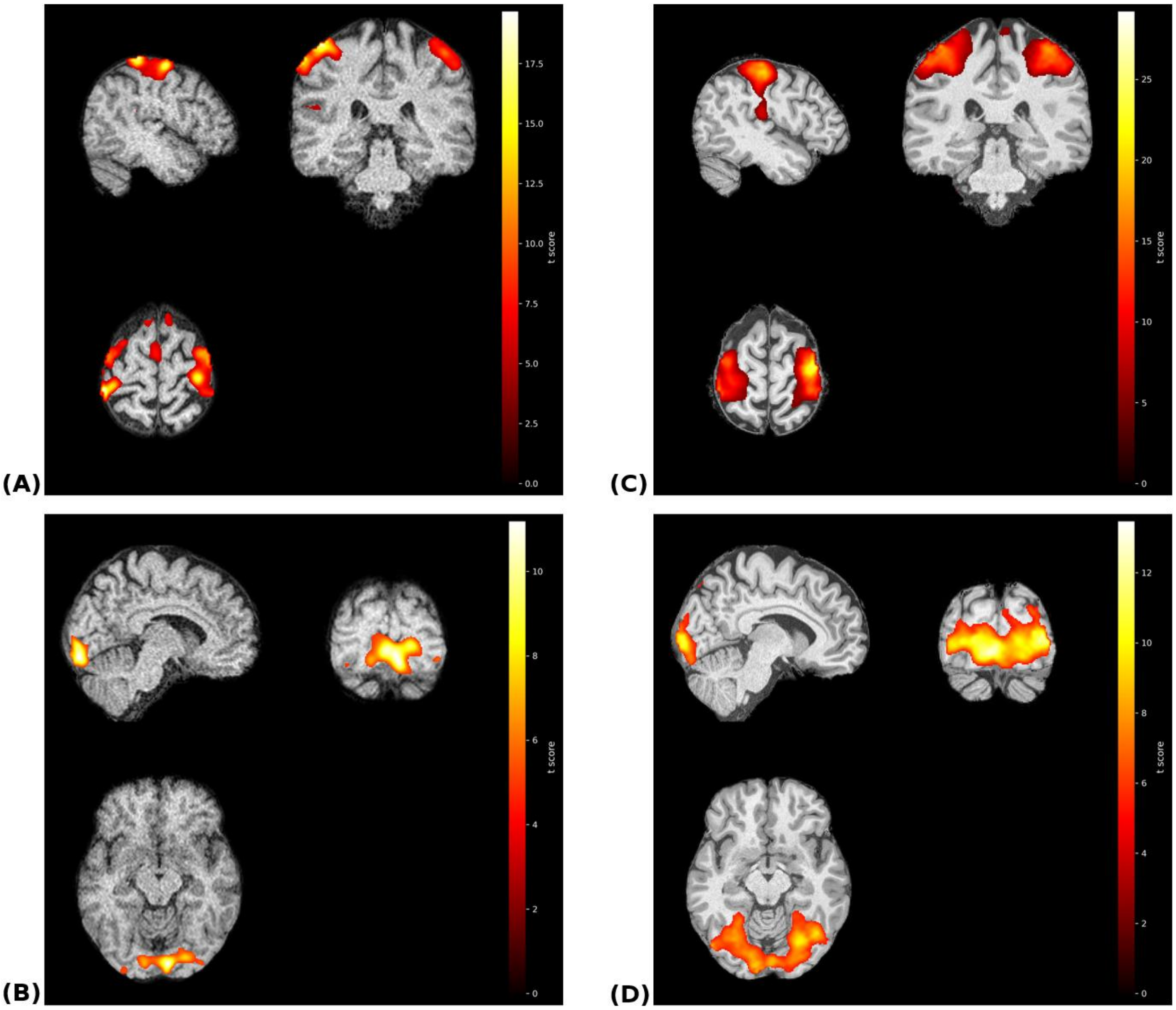

**Supporting Information Figure S6.** Orthogonal views of fMRI BOLD activation from the same participant scanned at 0.55T (10 min) and 3T (5 min). Motor task: (a) 0.55T and (c) 3T. Visual task: (b) 0.55T and (d) 3T. Panels (a,b) are overlaid on the T1-weighted anatomy acquired at 0.55T, while panels (c,d) are overlaid on the T1-weighted anatomy acquired at 3T. All panels show matched slice locations for qualitative comparison.

| ID | | Voxels | T$peak$ | x / y / z |
|---|---|---|---|---|
| VOL1 | Run1 | 1398 | 15.83 | -44 / -10 / 58 |
| | | 1396 | 13.12 | 34 / -18 / 70 |
| | Run2 | 822 | 13.96 | 34 / -26 / 62 |
| | | 882 | 12.28 | -42 / -12 / 62 |
| VOL2 | Run1 | 630 | 12.32 | 28 / -16 / 76 |
| | | 340 | 10.64 | -32 / -20 / 66 |
| | Run2 | 253 | 9.57 | 32 / -18 / 72 |
| | | 5 | 6.03 | -32 / -4 / 72 |
| VOL3 | Run1 | 532 | 11.84 | -44 / -10 / 56 |
| | | 302 | 8.13 | 36 / -22 / 66 |
| | Run2 | 251 | 11.41 | -44 / -10 / 56 |
| | | 51 | 8.05 | -34 / -30 / 72 |
| VOL4 | Run1 | 11 | 6.04 | -36 / -8 / 66 |
| | | 5 | 5.92 | 40 / -34 / 68 |
| | Run2 | 29 | 7.42 | -32 / -32 / 64 |
| | | 41 | 6.78 | 30 / -32 / 72 |
| VOL5 | Run1 | 106 | 7.27 | 36 / -34 / 68 |
| | | 17 | 6.39 | -26 / -20 / 70 |
| | Run2 | 142 | 9.45 | -32 / -18 / 68 |
| | | 123 | 8.81 | 36/ -32 / 68 |

**Supporting Information Table S1. Within-session consistency** of task-based fMRI activations for the bilateral finger-tapping task across two consecutive runs per participant at 0.55T. For each run, significant activation clusters are reported with their size, the t-score of the peak activation, FWE-corrected p-value ($p < 0.05$), and the MNI coordinates (X, Y, Z) of the peak voxel. These results allow comparison of activation patterns across repeated runs within a session, illustrating the consistency of motor responses.

| ID | Voxels | Tpeak | x / y / z |
|---|---|---|---|
| VOL1 | 2070 | 19.19 | -44 / -10 / 58 |
| | 2107 | 16.81 | 34 / -26 / 62 |
| VOL2 | 861 | 15.11 | 32 / -16 / 72 |
| | 498 | 10.77 | -32 / -20 / 66 |
| VOL3 | 1135 | 16.35 | -44 / -10 / 56 |
| | 518 | 9.74 | 38 / -20 / 64 |
| VOL4 | 150 | 8.95 | -32 / -32 / 64 |
| | 126 | 8.41 | 40 / -34 / 68 |
| VOL5 | 256 | 11.06 | 36 / -34 / 68 |
| | 249 | 10.36 | -26 / -18 / 70 |

**Supporting Information Table S2.** Peak activation statistics for the bilateral finger-tapping task at 0.55T (concatenated 10-min runs). For each participant, significant activation cluster(s) are reported for both hemispheres with cluster size, peak t-score, FWE-corrected p-value ($p < 0.001$), and MNI coordinates (X, Y, Z) of the peak activation.

| ID | | Voxels | T*peak* | x / y / z |
|---|---|---|---|---|
| VOL1 | Run1 | 275 | 8.71 | 9 / -86 / -17 |
| | Run2 | 31 | 7.08 | 6 / -89 / -8 |
| VOL2 | Run1 | 7 | 5.79 | 22 / -96 / -12 |
| | | 7 | 5.7 | -14 / -92 / -2 |
| | Run2 | 197 | 8.1 | -12 / -90 / 0 |
| | | 46 | 6.45 | 34 / -88 / -10 |
| VOL3 | Run1 | 79 | 7.32 | 8 / -82 / 8 |
| | | 39 | 7.21 | -26 / -92 / 20 |
| | Run2 | 90 | 7.71 | 10 / -84 / 8 |
| | | 9 | 5.55 | -12 / -98 / 6 |
| VOL4 | Run1 | 21 | 6.38 | -46 / -76 / -6 |
| | | 8 | 5.83 | -4 / -102 / 8 |
| | Run2 | 6 | 6.11 | 36 / -70 / -6 |
| | | 2 | 5.53 | -18 / -98 / 12 |
| VOL5 | Run1 | 29 | 9.16 | 24 / -70 / -12 |
| | | 18 | 6.52 | 0 / -94 / 6 |
| | Run2 | 51 | 7.62 | 6 / -84 / -2 |
| | | 25 | 7.22 | 0 / -96 / 6 |
| VOL6 | Run1 | 30 | 6.91 | -18 / -102 / 12 |
| | | 60 | 6.49 | -10 / -88 / 6 |
| | Run2 | 1 | 5.31 | 8 / -90 / 10 |

**Supporting Information Table S3. Within-session consistency** of task-based fMRI activations for the visual checkerboard task across two consecutive runs per participant at 0.55T. For each run, significant activation cluster(s) are reported with their size, the t-score of the peak activation, FWE-corrected p-value ($p < 0.05$), and the MNI coordinates (X, Y, Z) of the peak voxel. These results allow comparison of activation patterns across repeated runs within a session, illustrating the consistency of visual responses.

| ID | Voxels | $T_{peak}$ | x / y / z |
|---|---|---|---|
| VOL1 | 302 | 11.03 | 6 / -89 / -8 |
| VOL2 | 633 | 9.47 | -14 / -90 / -2 |
| | 66 | 6.73 | 34 / -90 / -10 |
| VOL3 | 340 | 9.85 | 8 / -84 / 8 |
| | 166 | 8.19 | -16 / -96 / 6 |
| VOL4 | 63 | 6.75 | -4 / -102 / 8 |
| | 94 | 6.15 | 30 / -86 / 16 |
| VOL5 | 42 | 10.75 | 24 / -70 / -12 |
| | 204 | 9.35 | 0 / -96 / 6 |
| VOL6 | 59 | 7.62 | 12 / -100 / 14 |
| | 43 | 7.03 | -18 / -102 / 12 |

**Supporting Information Table S4.** Visual task activation results at 0.55T (concatenated 10-min runs). For each run, the table lists the significant activation cluster(s) with their size, along with the t-score, FWE-corrected p-value ($p < 0.001$), and MNI coordinates (X, Y, Z) of the peak activations.

| ID | One Run | | | Two Runs | | |
|---|---|---|---|---|---|---|
| | Mean (μ) | SD (σ) | TV | Mean (μ) | SD (σ) | TV |
| VOL1 | 0.177 | 0.074 | 6.381 | 0.063 | 0.046 | 4.402 |
| VOL2 | 0.147 | 0.085 | 7.372 | 0.074 | 0.042 | 3.822 |
| VOL3 | 0.207 | 0.145 | 13.239 | 0.090 | 0.067 | 5.890 |
| VOL4 | 0.185 | 0.071 | 6.318 | 0.097 | 0.048 | 4.150 |
| VOL5 | 0.136 | 0.108 | 8.272 | 0.059 | 0.054 | 5.513 |
| VOL6 | 0.161 | 0.083 | 7.679 | 0.052 | 0.045 | 4.699 |

**Supporting Information Table S5.** Per-participant non-target ROI noise metrics for the visual task at 0.55T. Values for mean non-target correlation (μ), dispersion (σ), and total variation (TV) are reported for single 5-minute runs versus concatenated 10-minute runs (two 5-min runs). Metrics were derived from Pearson correlation coefficients (r) averaged over three non-target ROIs (200 voxels each) per participant.